# Faraday effect in bi-periodic photonic-magnonic crystals

Yuliya S. Dadoenkova[1,2,3], Nataliya N. Dadoenkova[2,3], Igor L. Lyubchanskii[3], Jarosław W. Kłos[4], and Maciej Krawczyk[4]

[1]Institute of Electronic and Information Systems, Novgorod State University, Veliky Novgorod, 173003, Russian Federation
[2]Technological Research Institute, Ulyanovsk State University, Ulyanovsk, 432017, Russian Federation
[3]Donetsk Physical and Technical Institute of the National Academy of Sciences of Ukraine, Donetsk, 83114, Ukraine
[4]Faculty of Physics, Adam Mickiewicz University in Poznań, Poznań, 61-614, Poland

**We present a theoretical investigation of the polarization plane rotation at light transmission - Faraday effect, through one-dimensional multilayered magneto-photonic systems consisting of periodically distributed magnetic and dielectric layers. We calculate Faraday rotation spectra of photonic-magnonic crystals, where cell (or supercell) is composed of magnetic layer and dielectric layer (or section of dielectric photonic crystal). We found that the Faraday rotation of *p*-polarized incident light is increasing in the transmission band with the number of magnetic supercells. The increase of Faraday rotation is observed also in vicinity of the band-gap modes localized in magnetic layers but the maximal polarization plane rotation angles are reached at minimal transmittivity. We show that presence of linear magneto-electric interaction in the magnetic layers leads to significant increase of the Faraday rotation angles of *s*-polarized incident light in the vicinity of the fine-structured modes inside the photonic-band-gap.**

## I. INTRODUCTION

THE magneto-optical interactions in photonics are interesting because of the application in non-reciprocal devices such as optical isolators and circulators [1] which are used for the routing of optical signals. These devices, based on the Faraday effect, or polarization plane rotation of light transmitted through magnetic medium, are necessary for the construction of optical systems for quantum metrology. The Faraday rotation (FR) can be also used for sensing of magnetic state [2] or even for chemical detection techniques [3]. The gaining of larger rotation angles of polarization plane in optical devices or achieving the sufficient sensitivity of detection based on the FR effect is however challenging in miniaturized systems. Therefore the search for materials [4] and structures [5] enhancing the FR is strongly needed. The one possible solution is to look for magneto-photonic composites in which we can expect the larger FR for the modes localized in magnetic subsystem [5], see also review papers [6], [7].

In our studies, we consider the magnetic photonic crystals (MPC) of double periodicity where magnetic layers are separated by the section of dielectric photonic crystal (DPC), known also as photonic-magnonic crystals [8]-[10]. In this bi-periodic structure, we investigate the light transmission in the vicinity of the modes in the frequency ranges, corresponding both to the frequency gaps and bands of infinite DPC. We show the enhanced FR in reference to conventional magneto-photonic structures.

The paper is organized as follows. In Section II we present geometry of the photonic structures and provide analytical equations for calculation of the FR angles. In Section III, we show and discuss the numerical calculations of the polarization plane rotation in different magneto-photonic structures. In Section IV, the Conclusions, we summarize the obtained results.

## II. DESCRIPTION OF THE SYSTEM

We investigate the 1D bi-periodic magneto-photonic systems consisting of magnetic layers *M* of thickness $d_M$ separated by non-magnetic dielectric spacers composed of layers *A* and *B* of thicknesses $d_A$ and $d_B$, respectively, as schematically illustrated in Fig. 1. First, we investigate simple structure with *ABA* three-layer of thickness $2d_A + d_B$ placed between the two layers *M* (structure *M (ABA) M*, Fig. 1(a)). Then, we consider the system with the sections of DPC: $(AB)^N A$ (of thickness $d_d = N(d_A + d_B) + d_A$), separated by magnetic layers *M*. The supercell $[M (AB)^N A]$ of this bi-periodic MPC is repeated *K* times, and the whole structure $[M (AB)^N A]^K M$ is completed by the additional magnetic layer M (see Fig. 1(b)). Finally (see Fig.1(c)), we investigate the reference structure where the sections of the DPC are replaced by homogeneous layer *A* or *B* of the same thickness $d_d$ as each DPC section in the structure presented in Fig.1(b)). We took into account two possible MPC structures: $[M A]^K M$ and $[M B]^K M$.

We assume that the electromagnetic wave of angular frequency $\omega$ is incident under angle $\theta$ from vacuum (with *xz*-plane being the incidence plane, see Fig. 1). The magnetic layers are magnetically saturated with the magnetization vector **M** lying in the incidence plane and parallel to the interfaces. This geometry corresponds to the longitudinal magneto-optical configuration [11].

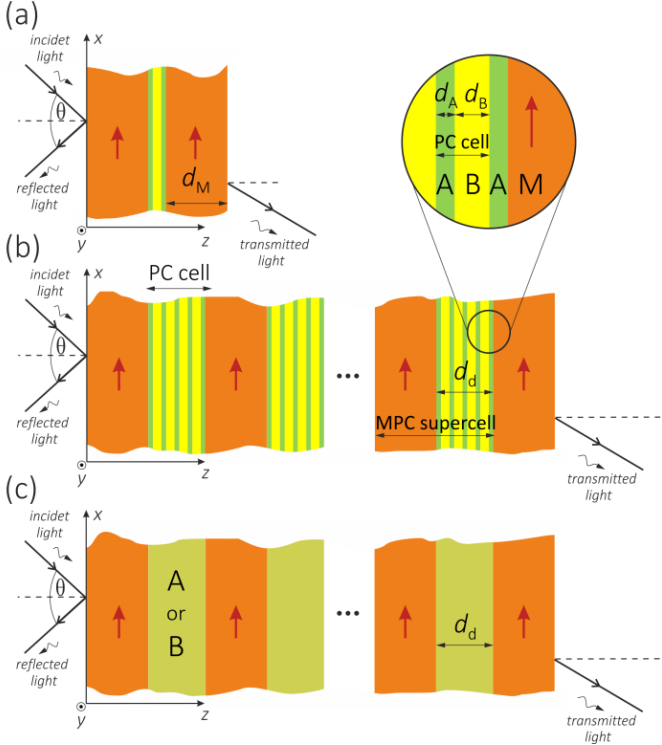

**FIG. 1.** Scheme of the magneto-photonic structures consisting of magnetic layers $M$ (thickness $d_M$) and dielectric layers $A$ and $B$ (thicknesses $d_A$ and $d_B$, respectively): (a) $M(ABA)M$; (b) $[M\,(AB)^N A]^K\,M$; (c) $[M\,A]^K\,M$ or $[M\,B]^K\,M$, where $N$ and $K$ denote the number of repetition of DPC cell and MPC supercell, respectively. Thickness of the non-magnetic spacers between the magnetic layers in (b) and (c) is $d_d$. The layers are infinitely extended in $x$- and $y$-directions. Red arrows show the direction of magnetization $\mathbf{M}$ in the magnetic layers, and $\theta$ is incidence angle of light.

The electric displacement vector $\mathbf{D}^{(C)}$ and the magnetic induction $\mathbf{B}^{(C)}$ in each layer $C$ ($C = A, B, M$) of the system are connected with the electric field $\mathbf{E}^{(C)}$ and the magnetic field $\mathbf{H}^{(C)}$ of the electromagnetic wave *via* constitutive material relations as follows:

$$D_i^{(C)} = \varepsilon_0 \varepsilon_{ij}^{(C)} E_j^{(C)},$$
$$B_i^{(C)} = \mu_0 \mu_{ij}^{(C)} H_j^{(C)}, \qquad (1)$$

where $\varepsilon_0$ and $\mu_0$ are the vacuum permittivity and permeability, and $\hat{\varepsilon}^{(C)}$ and $\hat{\mu}^{(C)}$ are tensors of dielectric permittivity and magnetic permeability of the corresponding medium $C$. In the isotropic $A$ and $B$ layers, $\varepsilon_{ij}^{(A,B)} = \varepsilon^{(A,B)} \delta_{ij}$ and $\mu_{ij}^{(A,B)} = \mu^{(A,B)} \delta_{ij}$, with $\delta_{ij}$ being the Kronecker symbol. Taking into account bigyrotropic properties of the magnetic medium, tensors $\hat{\varepsilon}^{(M)}$ and $\hat{\mu}^{(M)}$ in the linear magneto-optical approximation have the following form [11]:

$$\hat{\varepsilon}^{(M)} = \begin{pmatrix} \varepsilon^{(M)} & 0 & 0 \\ 0 & \varepsilon^{(M)} & i\varepsilon' \\ 0 & -i\varepsilon' & \varepsilon^{(M)} \end{pmatrix}, \qquad (2a)$$

$$\hat{\mu}^{(M)} = \begin{pmatrix} \mu^{(M)} & 0 & 0 \\ 0 & \mu^{(M)} & i\mu' \\ 0 & -i\mu' & \mu^{(M)} \end{pmatrix}. \qquad (2b)$$

We use $4 \times 4$ matrix method [12] to calculate the transmission matrix $\hat{T}$ whose components connect amplitudes of the transmitted wave $E_{p,s}^{(t)}$ to those of the incident one $E_{p,s}^{(i)}$ as:

$$\begin{pmatrix} E_p^{(t)} \\ E_s^{(t)} \end{pmatrix} = \begin{pmatrix} T_{pp} & T_{ps} \\ T_{sp} & T_{ss} \end{pmatrix} \begin{pmatrix} E_p^{(i)} \\ E_s^{(i)} \end{pmatrix}, \quad (3)$$

where the subscripts $p$ and $s$ refer to $p$- and $s$-polarizations. The off-diagonal components of the transmission matrix $\hat{T}$ correspond to the cross-polarized contribution to the transmission due to bigyrotropic properties of the magnetic layers.

The rotation of the polarization plane of the transmitted light, or FR angle, can be calculated in terms of the ratio $\eta = E_s^{(t)}/E_p^{(t)}$ as [13]:

$$\phi = \frac{1}{2} \arctan\left( \frac{2\,\mathrm{Re}(\eta)}{1-|\eta|^2} \right). \quad (4)$$

### III. Results of Numerical Calculations

For the numerical calculations, we consider the magnetic layers to be of yttrium-iron garnet (YIG) $Y_3Fe_5O_{12}$, which is transparent in near-infrared regime and possesses bigyrotropic properties. As layers $A$ and $B$ we take titanium oxide $TiO_2$ and silicon oxide $SiO_2$, respectively. We take into account the dielectric permittivity dispersion of all constituent media of the considered photonic structures. In the optical and near infra-red regimes the permittivity dispersion of $TiO_2$, $SiO_2$ and YIG has the following form [14], [15]:

$$\varepsilon^{(C)}(\lambda) = f_0^{(C)} + \sum_{i=1}^{3} \frac{f_i^{(C)} \lambda^2}{\lambda^2 - \left(\lambda_i^{(C)}\right)^2}, \quad C = A, B, M, \quad (5)$$

where the light wavelength $\lambda$ is in microns and the Sellmeier coefficients $f_i^{(C)}$ and $\lambda_i^{(C)}$ ($i = 1, 2, 3$) are gathered in Table I.

The off-diagonal material tensor elements of YIG are $\varepsilon' = -2.47 \cdot 10^{-4}$ and $\mu' = 8.76 \cdot 10^{-5}$ [16], and for the considered frequency regime $\mu_m = 1$ [17]. The thicknesses of the layers are taken to be $d_M = 700$ nm, $d_A = 190$ nm, and $d_B = 285$ nm, which provides a photonic band gap in the transmittivity spectra of the MPCs under consideration [8], [9].

TABLE I
SELLMEIER COEFFICIENTS OF THE MATERIALS

| Material | Sellmeier coefficients |
|---|---|
| $TiO_2$ [14] | $f_0^{(A)} = 5.913$, $f_1^{(A)} = 0.2441$, $f_2^{(A)} = 0$, $f_3^{(A)} = 0$, $\lambda_1^{(A)} = 0.0803\,\mu m$, $\lambda_2^{(A)} = 0$, $\lambda_3^{(A)} = 0$. |
| $SiO_2$ [14] | $f_0^{(B)} = 1$, $f_1^{(B)} = 0.6961663$, $f_2^{(B)} = 0.4079426$, $f_3^{(B)} = 0.8974794$, $\lambda_1^{(B)} = 0.0684043\,\mu m$, $\lambda_2^{(B)} = 0.1162414\,\mu m$, $\lambda_3^{(B)} = 9.896162\,\mu m$. |
| $Y_3Fe_5O_{12}$ [15] | $f_0^{(M)} = 1$, $f_1^{(M)} = 3.739$, $f_2^{(M)} = 0.79$, $f_3^{(M)} = 0$, $\lambda_1^{(M)} = 0.28\,\mu m$, $\lambda_2^{(M)} = 10.00\,\mu m$, $\lambda_3^{(M)} = 0$. |

#### A. Faraday rotation spectra

Fig. 2 presents FR angles $\phi_p$ of $p$-polarized (left panel) and $\phi_s$ of $s$-polarized (right panel) incident electromagnetic wave as functions of its incidence angle $\theta$ and angular frequency $\omega$.

In the structure $M\,(ABA)\,M$ both $\phi_p$ and $\phi_s$ possess negative values overall less than $0.1°$ (Fig. 2(a)). The FR angles in the structure $M\,(AB)^4 A\,M$ with 4 dielectric unit cells are illustrated in Fig. 2(b). Addition of dielectric unit cells ($AB$) between the magnetic layers leads to forming a photonic bad gap (PBG) in the transmittivity spectra of the structure, and narrow-frequency defect modes of high transmittivity appear inside the PBG due to magnetic layers on both sides of the DPC which act as defect layers. In Fig. 2(b) the positions of these defect modes and PBG edges where the transmittivity is maximal for each frequency are shown with the dotted white lines. One can see that the FR is minimal when the transmittivity is maximal. However, $\phi_s$ increases up to $-0.6°$ at large incidence angles, in contrast to $\phi_p$ which only slightly changes in comparison to that for the structure $M(ABA)M$ (see Figs. 2(a) and 2(b)).

With increase of the number $K$ of the magnetic supercells $M(AB)^4A$ each inside-PBG mode becomes narrower and splits into set of sub-peaks [8]-[10]. The FR angles inside the PBGs decrease [see Figs. 2(c) and 2(d) for the structures $[M(AB)^4A]^2M$ with $K = 2$ magnetic super-cell and $[M(AB)^4A]^5M$ with $K = 5$ magnetic super-cell, respectively], but in these cases $\phi_p$ and $\phi_s$ can be as negative, as positive. The behavior of the FR angles in the vicinity of the fine-structured inside-PBG modes (depicted with the black circles in Fig. 2(d)) will be discussed in Section B.

In bi-periodic MPCs, only $\phi_s$ remains a tendency to increase (in absolute value) at large $\theta$. It should be noted that polarization plane rotation of $s$-polarized incoming light transmitted through $[M(AB)^4A]^5M$ structure can reach ±45° at $\theta > 30°$ (not shown in Fig. 2), but these values of $\phi_s$ correspond to almost zero transmittivity. The maximal FR (about 0.1°–0.2°) in the bi-periodic MPCs is observed in the transmission band for $p$-polarized light, as one can see from the left panels in Figs. 2(c) and 2(d), where the transmission region lies right to the high-frequency dotted line. Note that the corresponding FR angles in the same frequency region for the structures without a double periodicity are minimal, as one can see comparing Figs. 2(a), 2(b) and 2(c), 2(d).

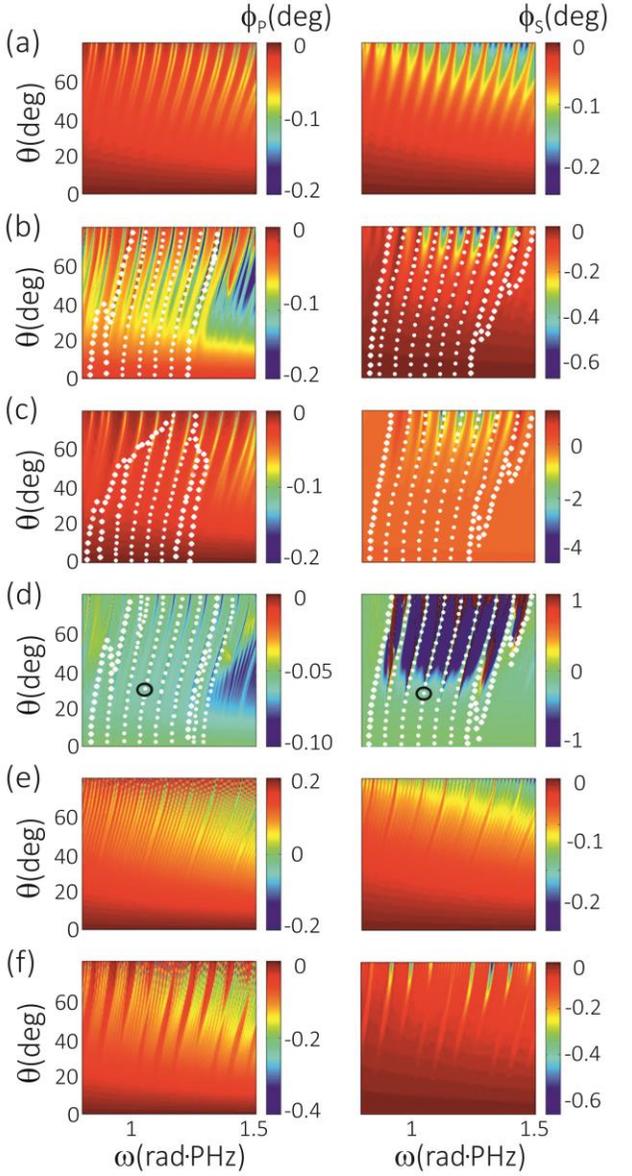

**FIG. 2.** Color maps of the evolution of the FR angles $\phi_p$ (left panel) and $\phi_s$ (right panel) of $p$- and $s$-polarized incident light, respectively, with frequency $\omega$ and incidence angle $\theta$ for the photonic structures: (a) $M(ABA)M$; (b) $M(AB)^4AM$; (c) $[M(AB)^4A]^2M$; (d) $[M(AB)^4A]^5M$; (e) $[M(A)]^5M$ with $d_A = d_d$; (f) $[MB]^5M$ with $d_B = d_d$. The white dots in (b), (c), and (d) depict the positions of the PBG edges and inside-PBG modes corresponding to the highest transmittivity values. The black circles in (d) show the spectral range detailed in Fig. 3.

For the structures with homogeneous dielectric spacer, $[M\,A]^5\,M$ and $[M\,B]^5\,M$ with $d_{A,B} = d_d$ (Figs. 2(e) and 2(f), respectively) no PBG occurs at the frequency range under consideration for the chosen thicknesses $d_d$ and $d_M$. The FR is negative and do not exceed $-1.5°$ at high $\theta$ for s-polarized light.

It should be noted that reversal of the magnetization of all magnetic layers leads to reversal of both FR angles $\phi_p$ and $\phi_s$.

*B. Influence of linear magneto-electric interaction on Faraday rotation angles*

It is a well-established fact that that a specific class of magnetic materials possesses spontaneous magneto-electric properties [18]. The magneto-electric effect, which results in the induction of a magnetization by an electric field or that of a dielectric polarization by a magnetic field, was observed in many systems, including garnets [19]. Thin epitaxial films of YIG can exhibit linear magneto-electric effect [19]. It was shown that this effect increases the cross-polarized contribution to the light reflected from a magneto-electric/dielectric bilayer [20], [21] and thus can increase the polarization plane rotation [22]. Thus in this subsection we estimate the possible impact of linear magneto-electric interaction on Faraday effect.

Taking into account the linear magneto-electric interaction, the material equations in the magnetic layers $M$ write [18]:

$$\begin{aligned} D_i^{(M)} &= \varepsilon_0 \varepsilon_{ij}^{(M)} E_j^{(M)} + \alpha_{ij} H_j^{(M)}, \\ B_i^{(M)} &= \mu_0 \mu_{ij}^{(M)} H_j^{(M)} + \alpha_{ij} E_j^{(M)}, \end{aligned} \quad (6)$$

where $\alpha_{ij}$ are the elements of the linear magneto-electric tensor of the magnetic medium, which is diagonal ($\alpha_{ij} = \alpha\,\delta_{ij}$) in crystals with a cubic symmetry [23]. It was shown that the magneto-electric constant in YIG can reach values $\alpha = 30$ ps·m$^{-1}$ [19]. The constitutive relations (6) after substitution to the Maxwell's equations give the eigenvectors of magnetic media. The magneto-electric constant a will modify the transmission properties of light through these wavevectors.

In Fig. 3 we show the influence of the magneto-electric interaction on transmittivity and FR angles of the light transmitted through a bi-periodic MPC of the structure $[M\,(AB)^4A]^5\,M$, focusing on the details of a single inside-PBG mode (see the black circles in Fig. 2(d)) when the angle of incidence is $\theta = 30°$. Solid lines correspond to the previously studied case when no magneto-electric interaction is present in the magnetic layers ($\alpha = 0$), and dotted lines refer to the case when the magneto-electric coupling is taken into account with $\alpha = 30$ ps·m$^{-1}$.

As was mentioned earlier, the inside-PBG modes in the transmittivity spectra of a bi-periodic MPC for all polarizations possess fine structure [8]-[10], and the number of the sub-peaks is related to the number of the magnetic super-cells of the structure. The sub-peaks number is the same in each mode in the $T_{pp}$ and $T_{ss}$ spectra [blue and pink lines in Fig. 3(a)], but it differs from that in $T_{ps}$ spectrum [Fig. 3(b)]. The frequency positions of the sub-peaks of $T_{pp}$ and $T_{ss}$ spectra are different but overlap with those in $T_{ps}$ spectrum [Figs. 3(a) and 3(b)]. The inside-PBG mode widths in $T_{pp}$ and $T_{ps}$ spectra ($\Delta\omega_{pp,\,ps} \approx 17$ THz) are larger than the width of the corresponding mode in $T_{ss}$ spectrum ($\Delta\omega_{ss} \approx 10$ THz).

The positions of the FR angle maxima do not correspond to position of the inside-PBG mode peaks in the spectra of the diagonal transmission coefficients, as follows from the comparison of blue and pink lines in Figs. 3(a) and 3(c). The origin of this behavior lies in the complex interplay between the diagonal and off-diagonal components of the transmission matrix, whose ratio defines the values of $\phi_{p,s}$ [see (4)]. However, $\phi_s$ is about one order of value larger than $\phi_p$ at the given $\theta = 30°$.

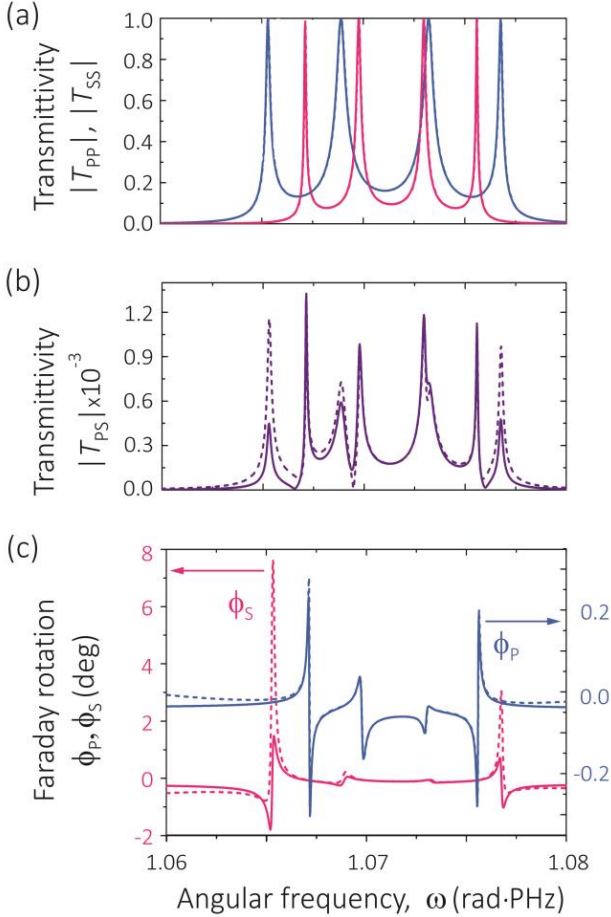

**FIG. 3**. Fine structure of (a) the inside-PBG modes in the transmission spectra $|T_{pp}|$ (blue lines) and $|T_{ss}|$ (pink lines); (b) $|T_{ps}|$; (c) FR angles $\phi_p$ (blue lines) and $\phi_s$ (pink lines) for the incidence angle $\theta = 30°$ and the structure $[M(AB)^4A]^5 M$ in the cases when the magneto-electric constant $\alpha = 0$ (solid lines) and $\alpha = 30$ ps·m$^{-1}$ (dotted lines).

The linear magneto-electric interaction almost does not change the diagonal components of the transmission matrix $T_{pp}$ and $T_{ss}$. On the contrary, as follows from the comparison of the solid and dotted lines in Fig. 3(b), the magneto-electric interaction results in increase the low- and high-frequency sub-peaks of the off-diagonal transmission coefficient $T_{ps}$.

The FR of $p$-polarized incident light almost does not change in presence of the magneto-electric interaction [solid and dotted blue lines in Fig. 3(c)], whereas for $\alpha = 30$ ps·m$^{-1}$ for $s$-polarized incoming light the positive low-frequency maximum of $\phi_s$ is five times as much as one for $\alpha = 0$, and its positive high-frequency maximum is about three times as much as without magneto-electric interaction [compare solid and dotted pink lines in Fig. 3(c)]. On the contrary, the negative maxima of $\phi_s$ decrease in presence of the magneto-electric interaction in the system.

## IV. CONCLUSIONS

We have investigated the influence of the spatial periodicity of the magnetic structure on Faraday rotation of near-infrared electromagnetic wave. We have shown that in bi-periodic magneto-photonic crystals the Faraday rotation of $p$-polarized incident light becomes feasible (about 0.2°) at the transmission band in comparison to that at light transmission through the structure with no double periodicity (where the polarization plane rotation is almost zero). At the vicinity of the inside-photonic-band-gap modes the Faraday rotation angles are also maximal (about 0.2°), but these values correspond to low transmittivity.

The magneto-electric coupling in the magnetic layers results in five-fold increase of the positive maxima of the polarization plane rotation angles of $s$-polarized incident light and decreases the negative ones, whereas the Faraday rotation of $p$-polarized light almost doesn't change in presence of the magneto-electric interaction.

## ACKNOWLEDGMENT

This work was supported in part by the Ministry of Education and Science of the Russian Federation under Projects No. 14.Z50.31.0015 and No. 3.7614.2017/ПЧ220; the Russian Science Foundation under Project No. 15-19-10036; the Ukrainian Fund for Fundamental Research under Project No. Ф71/73-2016; and the European Union's Horizon 2020 research and innovation programme under the Marie Skłodowska-Curie under Grant No. 644348.